# Current Poisson's ratio values of finite element models are too low to consider soft tissues nearly-incompressible: illustration on the human heel region

**Nolwenn Fougeron**[1], **Alessio Trebbi**[1], **Bethany Keenan**[2], **Yohan Payan**[1], **Gregory Chagnon**[1]

[1] Univ. Grenoble Alpes, CNRS, UMR 5525, VetAgro Sup, Grenoble INP, TIMC, 38000 Grenoble, France

[2] Cardiff School of Engineering, Cardiff University, Queens Buildings, 14-17 The Parade, Cardiff, CF24 3AA, UK

Corresponding author:

**Yohan Payan**

**TIMC Biomeca**

**Univ. Grenoble Alpes, CNRS, UMR 5525**

**Pavillon Taillefer, Allée des Alpes 38700 La Tronche**

E-mail: yohan.payan@univ-grenoble-alpes.fr

Keywords: Poisson ratio, Soft tissues, Incompressibility, Volume Changes, Strains

# Abstract

Finite element analysis of soft tissues is a well-developed method that allows estimation of mechanical quantities (e.g. stresses, strains). A constitutive law has to be used to characterise the individual tissues. This is complex as biological tissues are generally visco-hyperelastic, anisotropic, and heterogenous. A specific characteristic, their nearly incompressibility, was well reported in the literature, but very little effort has been made to compare volume variations computed by the simulations with *in vivo* measurements. In the present study, volume changes of the fat pad during controlled indentations of the human heel region were estimated from segmented medical images using digital volume correlation. Indentations were repeated with high and mild intensity normal and shear loads. The experiment was reproduced using finite element modelling with several values of Poisson's ratio for the fat pad, extracted from literature values (from 0.4500 to 0.4999). Estimated fat pad volume changes were compared to the measured ones to assess the best value of Poisson's ratio in each indentation case. The impact of the Poisson's ratio on the Jacobian of the deformation gradient and the volumetric strains was also computed. A single value of Poisson's ratio could not fit all the indentation cases. Estimated volume changes were between 0.9 % - 11.7 % with a Poisson's ratio from 0.4500 to 0.4999. The best fit was obtained with a 0.4900 Poisson's ratio except for the high normal load where a value of 0.4999 resulted in less error. In conclusion, special care should be taken when setting the Poisson's ratio as the resulting estimated deformations may become unrealistic when the value is far from incompressible materials.

*Word count: 248*

# 1. Introduction

Soft tissues are complex mechanical entities that connect, protect and support the human body (Holzapfel, 2000). They are composed of cells separated by the extracellular matrix and, at the microscopic scale, they can be described as a fibre-reinforced composite material constituted mainly by collagen and elastin proteins. The term "soft" refers to their high flexibility and their ability to undergo large deformations. They are heterogeneous and exhibit non-linear hyperelastic anisotropic behaviour (Fung, 1967). Furthermore, the response of soft tissues to external loads depends on the strain rate (Gennisson *et al.*, 2010). Due to their high water content, soft tissues are considered as nearly-incompressible meaning that there is little volume change when they are submitted to external loads.

Understanding the deformations of soft tissues is crucial in most clinical practices: tumour growth analysis (Iranmanesh and Nazari, 2017), tissue arrangement in the prosthetic socket (Moerman, Herr and Sengeh, 2016; Fougeron, Rohan, *et al.*, 2022), weight-bearing areas for pressure ulcer prevention (Ceelen, Stekelenburg, Mulders, *et al.*, 2008; Macron *et al.*, 2019; Peko, Barakat-Johnson and Gefen, 2020; Fougeron, Connesson, *et al.*, 2022). For that purpose, numerical methods, such as Finite Element (FE) analysis, have been developed for surgical planning, patient monitoring or orthopaedic device design. Constitutive laws were created to describe the intricate behaviour of soft tissues with material properties that could be tuned to fit subject-specific data. *In vivo,* studies have used imaging (Gefen *et al.*, 2001; Gennisson *et al.*, 2010), indentations (Zheng and Mak, 1996; Lin *et al.*, 2004; Sengeh, 2016), or both techniques (Affagard, Bensamoun and Feissel, 2014; Fougeron *et al.*, 2020) to quantify the material properties of the tissue. Indentation and inverse FE analysis could be applied to estimate material parameters for the soft tissues. Yet, these parameters may be biased by the simplifications of the indentation in the modelling process (Zhang, Zheng and Mak, 1997; Spears and Miller-Young, 2006). Besides the inherent difficulties associated with the process of material parameter identification, a limitation of most FE models is the lack of personalised data regarding the compressibility of the soft tissues. Soft tissues are mostly modelled nearly incompressible, whereas fluid exchanges occur in the human body (Swartz and Fleury, 2007). To what extent these fluid exchanges affect the nearly-incompressibility of soft tissues is still an open question. Mixed *pressure-displacement* formulation implemented in most FE software can be used to find a numerical solution when materials tend to be incompressible. In this case, the strain energy density is the combination of an isochoric part, that transcripts the transformation at constant volume, and the volumetric part which depends on the Jacobian of the deformation gradient and the initial bulk modulus of the materials to reflect the volume changes (Doll and Schweizerhof, 2000). The bulk modulus is a mechanical parameter that reflects the soft tissues' nearly-incompressibility. It can be expressed with the initial shear modulus or Young's modulus and Poisson's ratio; however, the latter is tedious to measure *in vivo*. It follows that the Poisson's ratio implemented in FE models extends to a wide range of values, going from 0.4500 to 0.4950 (Dickinson, Steer and Worsley, 2017; Keenan, Evans and Oomens, 2021). Some authors suggested considering the bulk-to-shear modulus ratio to quantify the level of incompressibility of hyperelastic materials (Love, 1892; Bonet and Wood, 2008). Suggested values of at least $10^3$ were recommended for isotropic incompressible materials. However, considering living tissues, and more particularly the arterial wall, (Toungara, Chagnon and Geindreau, 2012) have shown that for anisotropic hyperelastic polynomial models, this ratio should be superior to $10^6$, but data are still lacking for living tissues. This recommendation is not followed in the literature studies, and the impact of how the level of incompressibility of the soft tissues affects the models' results has been poorly investigated. Vannah et al. (Vannah and Childress, 1993) computed von Mises stresses and reaction force in a model of the contained residual limb of an above-knee amputated subject. Toungara et al. (Toungara, Chagnon and Geindreau, 2012) also investigated the first principal stress and the Jacobian of the deformation gradient of the arterial wall. Both authors concluded that changes in the Poisson ratio induced non-negligible changes in the computation of stresses. To the best of the authors' knowledge, no attempt has been made to evaluate this impact on soft tissue strains which are also an important mechanical quantity in many clinical fields. Pressure ulcers, for example, are well-known soft tissue injuries that are tedious to heal and negatively impact the physical but also mental health of the patients (Demarré *et al.*, 2015). This type of injury mainly occurs under bony prominences such as the sacrum and the heel regions and has been correlated to the shear strains in soft tissues (Ceelen, Stekelenburg, Loerakker, *et al.*, 2008). In addition, the level of incompressibility of living soft tissues is still required to be estimated from *in vivo* measurements.

This study aims to draw attention to the important impact of the Poisson'ratio on the mechanical response of soft tissues to external loading. The authors assumed that low Poisson's ratio, lower than 0.4900, cannot be representative of the nearly incompressibility of the soft tissues. In this study, Magnetic Resonance Imaging (MRI) were used to compute the changes in the volume of the soft tissues from Digital Volume Correlation (DVC). The foot was loaded with both normal and shear forces, with mild to high intensities, to analyze the resulting volume changes of the soft tissues in the fat pad. Indentations were reproduced in FE analysis of the heel region with several Poisson's ratios from 0.4500 to 0.4999. Volume changes, Jacobian, and volumetric strains in the fat pad were computed for all values of Poisson's ratio.

# 2. Material and methods

## 2.1. Experimental acquisition and processing

### 2.1.1. *MRI acquisitions*

Data were extracted from a previous work. For further information, the reader is invited to refer to the paper of Trebbi et al. 2021 (Trebbi *et al.*, 2021). A healthy volunteer (male, 40 years old) gave his informed consent to participate in the experimental part of a pilot study approved by an ethical committee (MammoBio MAP-VS pilot study N°ID RCB 2012-A00310-43, IRMaGe platform, Univ. Grenoble Alpes). A proton density MR was used to collect 512 consecutive 0.3 mm thick sagittal slices (Philips Achieva 3.0T dStream MRI system). Each slice had a field of view of 160.0 × 160.0 mm and a resolution of 512 × 428 for a total scanning time of 7 minutes. The right foot, in the supine position, was inserted into a compression device and fixed inside the foot casing. Five loading configurations, named from LOAD0 to LOAD4, were chosen to capture the non-linear mechanical properties of the soft tissues, and applied with the indenter. The complete protocol is presented in Figure 1. The displacements and reaction forces applied by the indenter, which consisted of a rigid plate, are detailed in Table 3. A double face tape was applied on the plate surface to allow the application of the shearing load without any slipping.

Figure 1 around here

### 2.1.2. *Image segmentation and soft tissue volume changes*

The MRI acquisitions were analysed using Amira (Amira Avizo 6.4, Thermos Fisher Scientific, Waltham, Massachusetts, United States). Only the acquisition in the undeformed configuration was manually segmented. The segmentation included the calcaneus, fat pad, Achilles tendon, muscle, skin and remaining bones and tissues. Due to the field of view of the MRI, only the calcaneus and the fat pad could be entirely segmented as shown in Figure 2. The volume of the fat pad, in the unloaded configuration, was calculated by multiplying the volume of one voxel by the number of voxels composing the segmented region. To compute the volume of the fat pad in the loaded configurations, the segmentation of the unloaded region was morphed to the loaded configuration using DVC through image registration. First, a rigid registration based on the calcaneus rotations and translations was performed to align all 3D images using Elastix libraries (Klein *et al.*, 2010) with a custom MATLAB code (MATLAB R2019b The MathWorks, Inc., Natick, Massachusetts, United States). This was followed by a non-rigid registration performed based on the computation of the normalised correlation coefficient, as the similarity measure between images, with the unloaded configuration set as the reference. Again, volumes of the fat pad in the loaded configurations were computed from the resulting segmentation by multiplying the number of voxels in the segmented volume by the volume of one voxel. Detailed procedure and accuracy analysis of this registration process have been proposed by (Trebbi *et al.*, 2022). The accuracy of the volumes was mainly characterized by the spatial resolution of the MRI 0.3 mm$^3$.

Figure 2 around here

## 2.2. FE ANALYSIS

### 2.2.1. *Geometry*

The segmented unloaded images were used to build the model of the foot which was composed of skin, fat pad, muscle and bones. The soft tissues of the upper part of the heel region and the bones were fused and were subsequently named the 'bony structure'. The bony structure was removed from the model as this component would be considered a rigid body in the analysis. No sliding was allowed at the tissues' interfaces. All components were meshed using PyAnsys (Kaszynski, 2021), used as an interface for ANSYS APDL (ANSYS 2020 R2 software, ANSYS Inc., Cannonsburg, PA, USA), with linear tetrahedron (SOLID285) with a mixed hydrostatic pressure and linear displacement formulation to avoid volumetric locking. A mesh convergence study was performed and eventually, a total of 296,408 degrees of freedom and 263,071 elements comprised the model (see Figure 2).

### 2.2.2. *Material properties*

Material parameters of all soft tissues were obtained from curve fitting of literature experimental data using MATLAB. The skin was modelled with the equation proposed by (Isihara, Hashitsume and

Tatibana, 1951) which is equivalent to the equation proposed by (Yeoh, 1990) with the parameter $C_{30}$ equals to zero, see equation 1. This model was fitted on the experimental data of (Ní Annaidh *et al.*, 2012) who did uniaxial tensile tests on skin samples. The fat pad layer was modelled with an Ogden 1st-order law (Ogden, 1972), see equation 2, and parameters were obtained from the compression tests performed (Miller-Young, Duncan and Baroud, 2002). Tendon and muscle tissues were both modelled with the equation proposed by Yeoh (1993), see equation 1. Tendon parameters were fitted according to the tensile test data published by Obuchowicz et al., (2019) whereas muscle parameters were fitted according to the tensile test data published by Gras et al. (2012). Material parameters of the soft tissues are summarised in Table 1.

Table 1 around here

Constitutive equations are provided below:

$$(1)\ W = \sum_{i=1}^{n} C_{i0}(\overline{I_1} - 3)^i + \sum_{l=1}^{n} \frac{1}{d}(J-1)^{2l}$$

$$(2)\ W = \frac{\mu_1}{\alpha_1}(\overline{\lambda_1}^{\alpha_1} + \overline{\lambda_2}^{\alpha_1} + \overline{\lambda_3}^{\alpha_1} - 3) + \frac{1}{d}(J-1)^2$$

$$(3)\ d = \frac{2}{\kappa}$$

with $W$ the strain energy density function, $C_{i0}$, $\mu_1$ and $\alpha_1$ are material parameters, $\overline{I_1}$ the first deviatoric invariant of the right Cauchy-Green deformation tensor, $\overline{\lambda_i}$ the principal stretches, $J$ the Jacobian, $d$ the incompressibility parameters and $\kappa$ the bulk modulus. Incompressibility parameters were computed for various Poisson's ratio values from the equation of (Mott, Dorgan and Roland, 2008) and are provided in Table 2. The values chosen were the most represented in the literature (Keenan, Evans and Oomens, 2021). A value of Poisson's ratio of 0.4999 was added to have a value above the incompressibility limit provided by (Bonet and Wood, 2008). In total, five models were constructed, one for each Poisson's ratio and the same Poisson's ratio was given to all tissues in each model.

Table 2 around here

### 2.2.3. *Boundary conditions*

No sliding was allowed between the skin and the plate due to the presence of tape during the experiments. The frontier with the bone was defined as rigid and was fixed in the simulation. Normal and shearing loads of LOAD1 to LOAD4 were applied to all models. Maximal values of normal and shear loads were equivalent to 15 % and 5 % of the subject body weight respectively. These values were chosen to capture the non-linear hyperelastic behaviour of the tissues without inducing pain in the subject. A step was defined to apply a vertical displacement of 10.0 mm to the plate to ensure contact with the soft tissues. Then, a second analysis step was created to apply the vertical force on the plate. When needed, a third step was added to remove the displacement constraints on the horizontal displacement of the plate and to apply the shearing load. Eventually, simulations were launched in quasi-static analysis using an implicit scheme. Loading details are provided in Table 3.

# 3. Results

## 3.1. Plate displacement

The displacement of the plate measured during the experiment was compared to the computed displacement of all models (detailed in Table 3). Differences are noticeable between the experimental measurements and the computed values. Vertical displacements are similar for all models and overestimate the indenter displacements. Horizontal displacements tend to be underestimated by the models. Except for LOAD 1 where errors were up to 5.1 mm, plate displacement errors were below 2.7 mm.

> Table 3 around here

## 3.2. Changes in soft tissues' volume

Experimental volume changes in the fat pad were computed from the absolute difference between the sum of the volume of the voxels in the deformed segmentations and the undeformed configuration and plotted in Figure 3. Numerical volume changes in the fat pad were computed from the absolute difference between the sum of the volume of the elements at the last step of the simulation and in the undeformed configuration. Volume changes were reported as a percentage of the volume computed in the undeformed configuration. Fat pad's volume changes were between 0.2 and 11.7 %. The range of volume changes was comprised in [0.9; 11.7] for LOAD 1, in [0.2; 1.9] for LOAD 2, in [0.4; 3.5] for LOAD 3 and in [0.2; 2.3] for LOAD 4. The higher the load intensity the higher the impact of the Poisson's ratio for both normal and shear loads. In this particular case, the Poisson's ratio $\nu = 0.4900$ provided the best fit of the fat pad's volume change except for LOAD 1 for which the Poisson's ratio $\nu = 0.4999$ gave the best results.

> Figure 3 around here

### 3.3. Jacobian and soft tissue compressibility

Considering a 1st-order Ogden constitutive equation, the ratio of the bulk-to-tangent shear modulus was computed from the fat pad's stress-strain curves. As illustrated by Figure 4, only the model with a Poisson's ratio $\nu$ = 0.4999 when the stretch ratio is above 0.66 (which is equivalent to a decrease of less than 33 % of the element in the compression direction) could be defined as incompressible with regard to the bulk-to-shear ratio threshold proposed by Bonet and Wood (Bonet and Wood, 2008). However, as expected, the median value of the Jacobian was 1.00 for all models and in high load cases, LOAD 1 and LOAD 3, the interquartile range decreased when the Poisson ratio increased. Figure 5 shows the boxplots of the soft tissues' Jacobian without outliers for clarity since the interquartile ranges were very low.

Figure 4 around here

Figure 5 around here

### 3.4. Volumetric strains

To highlight the impact of Poisson's ratio on some mechanical data of interest, the volumetric strains of the fat pad were computed for all models. Absolute values of mean and maximal values are provided in Table 4. The impact of the Poisson's ratio is very small at low loads, yet maximal volumetric strains decreased by 16 % and 18 % for LOAD 1 (high normal load) and LOAD 3 (high shear load) respectively. Evaluation of the computed strains was proposed in a previous study (Trebbi, Fougeron and Payan, 2023).

Table 4 around here

# 4. Discussion

The modelling of soft tissues in interacting with medical devices is an important field of research and has many applications in the clinical field. Yet, the numerical analysis of this interaction requires a correct definition of the soft tissues' response to external loads. The mechanical characterisation of soft tissues is challenging due to their highly complex behaviour under large displacement, which varies considerably per person. Many researchers have proposed *ex vivo* and *in vivo* methods for the estimation of the material properties of soft tissues. However, the bulk modulus has received little attention in the literature until now. The nearly-incompressibility of the soft tissues is accepted but there is no consensus on the corresponding bulk modulus, usually transcripted in terms of Poisson's ratio, to be implemented numerically. In this study, we wanted to highlight the impact of the bulk modulus on the volume changes of the fat pad of the heel region in FE models. These models were evaluated regarding strain prediction in a previous study (Trebbi, Fougeron and Payan, 2023). Soft tissues are mostly modelled nearly-incompressible, whereas fluid exchanges occur in the human body. The numerical predictions of volume changes were compared with experimental results obtained from DVC on MRI data. The material properties of the soft tissues were inferred from mechanical tests of the literature. To ease the interpretation and comparison with models of the literature, the results were expressed with the Poisson's ratio. The Poisson's ratio varied between $\nu = 0.4500$ and $\nu = 0.4950$ since these values are the most commonly found in models (Dickinson, Steer and Worsley, 2017; Keenan, Evans and Oomens, 2021). The value $\nu = 0.4999$ was added to have at least one bulk-to-shear modulus ratio above the incompressibility threshold described by Bonet and Wood (Bonet and Wood, 2008).

The results from this study showed that the values of the Poisson's ratio affected the volume changes of the fat pad since in all cases decreasing the Poisson's ratio induced an increase in the volume changes. A change of the Poisson's ratio from $\nu = 0.4500$ to $\nu = 0.4999$ induced a multiplication by a factor between 9 and 11 of the volume changes for all loads. This can be explained by the changes in the Jacobian. In the current study, this Jacobian was more spread for low values of the Poisson's ratio. As the Poisson's ratio increased the median of the Jacobian became closer to one, thus, as expected, tissues were less compressible (Doll and Schweizerhof, 2000). This impact was also visible in the values of the volumetric strains for which the increase of the Poisson's ratio led to a decrease of 16 % and 18 % of the maximal volumetric strain when high loads were applied to the heel region. Vannah et al. (Vannah and Childress, 1993) estimated the impact of the Poisson's ratio on the reaction force applied to the residual limb inside the prosthetic socket and on the soft tissues' von Mises stresses. Considering a linear elastic behaviour for the soft tissues, the authors showed that increasing the Poisson's ratio from $\nu$ = 0.4500 to $\nu$ = 0.4999 increased by 46.7 N of the reaction force. Von Mises stresses also increased from 0.12 to 0.66. Although a direct comparison cannot be made, the increase in stress observed by Vannah et al. agrees with the reduction in strain shown by the current study. Toungara et al. (Toungara, Chagnon and Geindreau, 2012) also study the impacts of the incompressibility hypothesis on the stress state of the arterial wall. The authors showed that increasing the Poisson's ratio from $\nu$ = 0.49000 to $\nu$ = 0.49999 resulted in an increase of the maximal first principal stress by 78 %. The authors also pointed out that even with high Poisson's ratios the incompressibility hypothesis is verified only until a given level of deformation depending on the constitutive equation. This is concurrent with the conclusion of the current study's bulk-to-tangent shear modulus ratio analysis. Only the model with the highest Poisson's ratio, $\nu$ = 0.4999, when the stretch ratio is above 0.66 could be considered incompressible based on the threshold provided by Bonet and Wood (Bonet and Wood, 2008). The impact of the Poisson's ratio on the previously detailed data was particularly visible at high normal and shear loads. This was also reported by Toungara et al. who showed that discrepancies between models with different Poisson's ratios were increasing with increasing first principal strains (Toungara, Chagnon and Geindreau, 2012).

Limitations of this study have to be reported. First, the material parameters of the soft tissues were estimated only on the isochoric part of the strain energy function and with generic mechanical tests. Consequently, the material parameters given in this study were not subject-specific and not affected by the changes in Poisson's ratio values. However, this study aimed to provide relative quantitative insights into the assumption of soft tissues' incompressibility. Thus subject-specific data were not required in this case. Constitutive behaviours defined from *in vivo* indentation tests were not used in the current study since these results may be biased by the modelling simplifications and the resulting identifiability parameter set may not be satisfying (Oddes and Solav, 2023). Further work will seek to address subject-specific material parameters, bulk modulus included, from inverse FE using MRI data such as tissue displacement and strain fields (Trebbi *et al.*, 2022) and soft-tissue volume changes. It is also worth noting that even though plate displacement errors are small, they still accounted for 12 % to 61 % of the experimental data, with the most important relative errors for low-intensity loads. These errors may also be explained by the fact that material parameters, bulk modulus included, were not subject-specific. In addition, Poisson's ratio values, used to compute the bulk modulus, were identical for all tissues whereas nothing guaranteed that one bulk modulus could be suitable for all tissues. Yet, compared to the MRI data the Poisson's ratio $\boldsymbol{\nu}$ = 0.4900 provided satisfying results in terms of the volume changes of the fat pad, except for high normal loads where a Poisson's ratio $\boldsymbol{\nu}$ = 0.4999 provided better results. This may be revealing of the weakness of solid monophysical FE models when it comes to the modelling of living soft tissues. Some studies have investigated the implementation of biphasic models (Miller, 1998; Sassaroli, O'Neill and Li, 2008; Sciumè *et al.*, 2014). Biphasic models can be used to model interstitial fluids in porous soft tissues but also account for the time-dependent response of the tissues. However, the rise in model complexity that requires additional material parameters, often tedious to assess in vivo, is an important obstacle to the spreading of this approach. The visco-hyperelastic material parameters were also neglected in this work. The indenter position was maintained for 5 seconds to limit the error due to the creep of the tissues (Trebbi *et al.*, 2021). Further work should be performed to assess the impact of the tissues' Poisson's ratio when the duration of the load is also considered.

Considering solid models of soft tissues, particular care should be taken when defining the material properties including the bulk modulus. Strains are an important mechanical quantity in many clinical applications such as the study of the onset of pressure ulcers. In addition, diseases or soft tissue injuries such as pressure ulcers may locally impact the incompressibility behaviour of the tissues. The current study focuses on the heel region, yet, many models of the sacrum, ischial tuberosities or face regions are also proposed in our group and the literature (Luboz *et al.*, 2014; Al-Dirini *et al.*, 2016; Levy and Gefen, 2017; Savonnet, Wang and Duprey, 2018; Macron *et al.*, 2019). It is worth noting that material parameters are specific to the region of interest (Zhang, Zheng and Mak, 1997); this is particularly important with regard to the Poisson's ratio. The impact of the bulk modulus on these strains was clear at high loads and it is not uncommon to have important loads applied to soft tissues in clinical fields. The bulk modulus could be validated using medical image analyses such as DVC or should at least be included in a sensitivity analysis to assess the confidence domain of the FE results.

# 5. Acknowledgment

The authors would like to acknowledge Urgo RID (21300, Chenôve, France) for their financial support in this study.

# 6. Conflict of interest

Nolwenn Fougeron has been financially supported by Urgo RID (21300, Chenôve, France) but owns no stock in this company and has thus no conflict of interests.

# 7. References

Affagard, J.-S., Bensamoun, S.F. and Feissel, P. (2014) 'Development of an Inverse Approach for the Characterization of In Vivo Mechanical Properties of the Lower Limb Muscles', *Journal of Biomechanical Engineering*, 136(11), p. 111012. Available at: https://doi.org/10.1115/1.4028490.

Al-Dirini, R.M.A. *et al.* (2016) 'Development and Validation of a High Anatomical Fidelity FE Model for the Buttock and Thigh of a Seated Individual', *Annals of Biomedical Engineering*, 44(9), pp. 2805–2816. Available at: https://doi.org/10.1007/s10439-016-1560-3.

Bonet, J. and Wood, R.D. (2008) *Nonlinear continuum mechanics for finite element analysis, 2nd edition*, *Nonlinear Continuum Mechanics for Finite Element Analysis, 2nd Edition*. Available at: https://doi.org/10.1017/CBO9780511755446.

Ceelen, K.K., Stekelenburg, A., Loerakker, S., *et al.* (2008) 'Compression-induced damage and internal tissue strains are related', *Journal of Biomechanics*, 41, pp. 3399–3404. Available at: https://doi.org/10.1016/j.jbiomech.2008.09.016.

Ceelen, K.K., Stekelenburg, A., Mulders, J.L.J., *et al.* (2008) 'Validation of a numerical model of skeletal muscle compression with MR tagging: A contribution to pressure ulcer research', *Journal of Biomechanical Engineering*, 130(6), pp. 1–8. Available at: https://doi.org/10.1115/1.2987877.

Demarré, L. *et al.* (2015) 'The cost of prevention and treatment of pressure ulcers: A systematic review', *International Journal of Nursing Studies*, 52(11), pp. 1754–1774. Available at: https://doi.org/10.1016/j.ijnurstu.2015.06.006.

Dickinson, A.S., Steer, J.W. and Worsley, P.R. (2017) 'Finite element analysis of the amputated lower limb: A systematic review and recommendations', *Medical Engineering and Physics*, 43, pp. 1–18. Available at: https://doi.org/10.1016/j.medengphy.2017.02.008.

Doll, S. and Schweizerhof, K. (2000) 'Volumetric Strain Energy', 67(March), pp. 17–21.

Fougeron, N. *et al.* (2020) 'Combining Freehand Ultrasound-Based Indentation and Inverse Finite Element Modeling for the Identification of Hyperelastic Material Properties of Thigh Soft Tissues', *Journal of Biomechanical Engineering*, 142(9). Available at: https://doi.org/10.1115/1.4046444.

Fougeron, N., Rohan, P., *et al.* (2022) 'Finite element analysis of the stump-ischial containment socket interaction: a technical note', *Medical Engineering & Physics*, 105.

Fougeron, N., Connesson, N., *et al.* (2022) 'New pressure ulcers dressings to alleviate human soft tissues: A finite element study', *Journal of Tissue Viability*, 31(3), pp. 506–513. Available at: https://doi.org/10.1016/j.jtv.2022.05.007.

Fung, Y.-C. (1967) 'Elasticity of soft tissues in simple elongation', *American Journal of Physiology-Legacy Content*, 213(6), pp. 1532–1544. Available at: https://doi.org/10.1152/ajplegacy.1967.213.6.1532.

Gefen, A. *et al.* (2001) 'Integration of plantar soft tissue stiffness measurements in routine MRI of the diabetic foot', *Clinical Biomechanics*, 16(10), pp. 921–925. Available at: https://doi.org/10.1016/S0268-0033(01)00074-2.

Gennisson, J.L. *et al.* (2010) 'Viscoelastic and anisotropic mechanical properties of in vivo muscle tissue assessed by supersonic shear imaging', *Ultrasound in Medicine and Biology*, 36(5), pp. 789–801. Available at: https://doi.org/10.1016/j.ultrasmedbio.2010.02.013.

Gras, L.L. *et al.* (2012) 'Hyper-elastic properties of the human sternocleidomastoideus muscle in tension', *Journal of the Mechanical Behavior of Biomedical Materials*, 15, pp. 131–140. Available at: https://doi.org/10.1016/j.jmbbm.2012.06.013.

Holzapfel, G.A. (2000) 'Biomechanics of Soft Tissue', *Computational Biomechanics* [Preprint]. Available at: https://doi.org/10.1109/CA.1999.781200.

Iranmanesh, F. and Nazari, M.A. (2017) 'Finite Element Modeling of Avascular Tumor Growth Using a Stress-Driven Model', *Journal of Biomechanical Engineering*, 139(8), pp. 1–10. Available at: https://doi.org/10.1115/1.4037038.

Isihara, A., Hashitsume, N. and Tatibana, M. (1951) 'Statistical theory of rubber-like elasticity. IV. (Two-dimensional stretching)', *The Journal of Chemical Physics*, 19(12), pp. 1508–1512. Available at: https://doi.org/10.1063/1.1748111.

Kaszynski, A. (2021) 'pyansys: Python Interface to MAPDL and Associated Binary and ASCII Files'. Zenodo. Available at: https://doi.org/https://doi.org/10.5281/zenodo.572600.

Keenan, B.E., Evans, S.L. and Oomens, C.W.J. (2021) 'A review of foot finite element modelling for pressure ulcer prevention in bedrest: Current perspectives and future recommendations', *Journal of Tissue Viability* [Preprint]. Available at: https://doi.org/10.1016/j.jtv.2021.06.004.

Klein, S. *et al.* (2010) 'Elastix: A toolbox for intensity-based medical image registration', *IEEE Transactions on Medical Imaging* [Preprint]. Available at: https://doi.org/10.1109/TMI.2009.2035616.

Levy, A. and Gefen, A. (2017) 'Assessment of the Biomechanical Effects of Prophylactic Sacral Dressings on Tissue Loads: A Computational Modeling Analysis', *Ostomy Wound Management*, 63(10), pp. 48–55. Available at: https://doi.org/10.25270/owm.10.4855.

Lin, F. *et al.* (2004) 'A subject-specific FEM model for evaluating buttock tissue response under sitting load', *Annual International Conference of the IEEE Engineering in Medicine and Biology - Proceedings*, 26 VII(C), pp. 5088–5091.

Love, A.M. (1892) *A Treatise on the Mathematical Theory of Elasticity*, *Nature*. Available at: https://doi.org/10.1038/105511a0.

Luboz, V. *et al.* (2014) 'Biomechanical modeling to prevent ischial pressure ulcers', *Journal of Biomechanics*, 47(10), pp. 2231–2236. Available at: https://doi.org/10.1016/j.jbiomech.2014.05.004p.

Macron, A. *et al.* (2019) 'Development and validation of a new methodology for the fast generation of patient-specific FE models of the buttock for pressure ulcer prevention', *Journal of Biomechanics* [Preprint].

Miller, K. (1998) 'Modelling soft tissue using biphasic theory - a word of caution', *Computer Methods in Biomechanics and Biomedical Engineering*, 1(3), pp. 261–263. Available at: https://doi.org/10.1080/01495739808936706.

Miller-Young, J.E., Duncan, N.A. and Baroud, G. (2002) 'Material properties of the human calcaneal fat pad in compression: Experiment and theory', *Journal of Biomechanics*, 35(12), pp. 1523–1531. Available at: https://doi.org/10.1016/S0021-9290(02)00090-8.

Moerman, K.M., Herr, H.M. and Sengeh, D.M. (2016) 'Automated and Data-driven Computational Design of Patient-Specific Biomechanical Interfaces', *Open Science Framework*, X, pp. 1–17. Available at: https://doi.org/10.17605/OSF.IO/G8H9N.

Mott, P.H., Dorgan, J.R. and Roland, C.M. (2008) 'The bulk modulus and Poisson's ratio of "incompressible" materials', *Journal of Sound and Vibration*, 312(4–5), pp. 572–575. Available at: https://doi.org/10.1016/j.jsv.2008.01.026.

Ní Annaidh, A. *et al.* (2012) 'Characterization of the anisotropic mechanical properties of excised human skin', *Journal of the Mechanical Behavior of Biomedical Materials*, 5(1), pp. 139–148. Available at: https://doi.org/10.1016/j.jmbbm.2011.08.016.

Obuchowicz, R. *et al.* (2019) 'Interfascicular matrix-mediated transverse deformation and sliding of discontinuous tendon subcomponents control the viscoelasticity and failure of tendons', *Journal of the Mechanical Behavior of Biomedical Materials*, 97, pp. 238–246. Available at: https://doi.org/10.1016/j.jmbbm.2019.05.027.

Oddes, Z. and Solav, D. (2023) 'Identifiability of soft tissue constitutive parameters from in-vivo macro-indentation', *Journal of the Mechanical Behavior of Biomedical Materials*, 140, p. 105708. Available at: https://doi.org/https://doi.org/10.1016/j.jmbbm.2023.105708.

Ogden, R.W. (1972) 'Large Deformation Isotropic Elasticity—On the Correlation of Theory and Experiment for Incompressible Rubberlike Solids', *Rubber Chemistry and Technology*, 46(2), pp. 398–416. Available at: https://doi.org/10.5254/1.3542910.

Peko, L., Barakat-Johnson, M. and Gefen, A. (2020) 'Protecting prone positioned patients from facial pressure ulcers using prophylactic dressings: A timely biomechanical analysis in the context of the COVID-19 pandemic', *International Wound Journal*, 17(6), pp. 1595–1606. Available at: https://doi.org/10.1111/IWJ.13435.

Sassaroli, E., O'Neill, B.E. and Li, K.C. (2008) 'Biphasic models of soft tissues for ultrasound applications', *Proceedings of Meetings on Acoustics*, 5. Available at: https://doi.org/10.1121/1.3046548.

Savonnet, L., Wang, X. and Duprey, S. (2018) 'Finite element models of the thigh-buttock complex for assessing static sitting discomfort and pressure sore risk: a literature review', *Computer Methods in*

*Biomechanics and Biomedical Engineering*, 21(4), pp. 379–388. Available at: https://doi.org/10.1080/10255842.2018.1466117.

Sciumè, G. *et al.* (2014) 'A two-phase model of plantar tissue: a step toward prediction of diabetic foot ulceration', *INTERNATIONAL JOURNAL FOR NUMERICAL METHODS IN BIOMEDICAL ENGINEERING*, 30, pp. 1153–1169. Available at: https://doi.org/10.1002/cnm.2650.

Sengeh, D.M. (2016) *The Use of a Novel Residuum Model to Design a Variable-Impedance Transtibial Prosthetic Socket*.

Spears, I.R. and Miller-Young, J.E. (2006) 'The effect of heel-pad thickness and loading protocol on measured heel-pad stiffness and a standardized protocol for inter-subject comparability', *Clinical Biomechanics* [Preprint]. Available at: https://doi.org/10.1016/j.clinbiomech.2005.09.017.

Swartz, M.A. and Fleury, M.E. (2007) 'Interstitial flow and its effects in soft tissues', *Annual Review of Biomedical Engineering*, 9, pp. 229–256. Available at: https://doi.org/10.1146/annurev.bioeng.9.060906.151850.

Toungara, M., Chagnon, G. and Geindreau, C. (2012) 'Numerical analysis of the wall stress in abdominal aortic aneurysm: Influence of the material model near-incompressibility', *Journal of Mechanics in Medicine and Biology*, 12(1), pp. 1–19. Available at: https://doi.org/10.1142/S0219519412004442.

Trebbi, A. *et al.* (2021) 'MR-compatible loading device for assessment of heel pad internal tissue displacements under shearing load.', *Medical Engineering and Physics*, 98(November), pp. 125–132. Available at: https://doi.org/10.1016/j.medengphy.2021.11.006.

Trebbi, A. *et al.* (2022) 'Medical Engineering and Physics MR-based quantitative measurement of human soft tissue internal strains for pressure ulcer prevention', *Medical Engineering and Physics* [Preprint].

Trebbi, A., Fougeron, N. and Payan, Y. (2023) 'Definition and evaluation of a finite element model of the human heel for diabetic foot ulcer prevention under shearing loads', *Medical Engineering & Physics*, 118, p. 104022. Available at: https://doi.org/10.1016/j.medengphy.2023.104022.

Vannah, W.M. and Childress, D.S. (1993) 'Modelling the mechanics of narrowly contained soft tissues: The effects of specification of Poisson's ratio', *Journal of Rehabilitation Research and Development*, 30(2), pp. 205–209.

Yeoh, O.H. (1990) 'Characterization of Elastic Properties of Carbon-Black-Filled Rubber Vulcanizates', *Rubber Chemistry and Technology*, pp. 792–805. Available at: https://doi.org/10.5254/1.3538289.

Yeoh, O.H. (1993) 'Some forms of the strain energy function for rubber', *Rubber Chemistry and Technology*, pp. 754–771. Available at: https://doi.org/10.5254/1.3538343.

Zhang, M., Zheng, Y.P. and Mak, A.F.T. (1997) 'Estimating the effective Young's modulus of soft tissues from indentation tests—nonlinear finite element analysis of effects of friction and large deformation', *Medical Engineering & Physics*, 19(6), pp. 512–517. Available at: https://doi.org/10.1016/S1350-4533(97)00017-9.

Zheng, Y.P. and Mak, A.F.T. (1996) 'An ultrasound indentation system for biomechanical properties assessment of soft tissues in-vivo', *IEEE Transactions on Biomedical Engineering*, 43(9), pp. 912–918. Available at: https://doi.org/10.1109/10.532125.

# 8. List of Figures

| | |
|---|---|
| Figure 1 | Experimental MRI acquisitions of the foot in various loading configurations (from Trebbi et al., [29]) |
| Figure 2 | MRI image of the unloaded configuration (a), LOAD 0, and the resulting segmented image (b). Segmented volumes were meshed to define the finite element model (c). |
| Figure 3 | Comparison of the soft tissues volume changes computed from experimental data and the finite element model for all Poisson ratios. |
| Figure 4 | Bulk-to-tangent shear modulus values for the fat pad for all Poisson's ratio values. (b) Zoomed bulk-to-tangent shear modulus curve. The red dot line represents the incompressibility threshold value proposed by Bonet and Wood. |
| Figure 5 | Fat pad's Jacobian in all loading conditions for all Poisson's ratio values. |

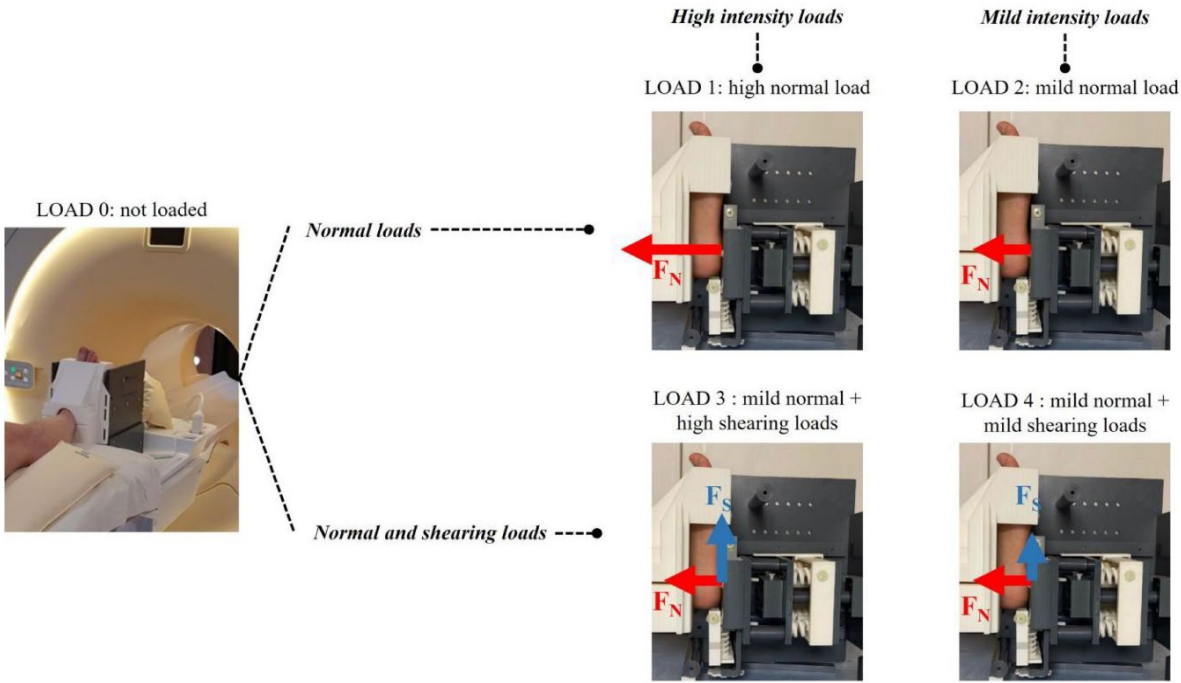

Figure 1

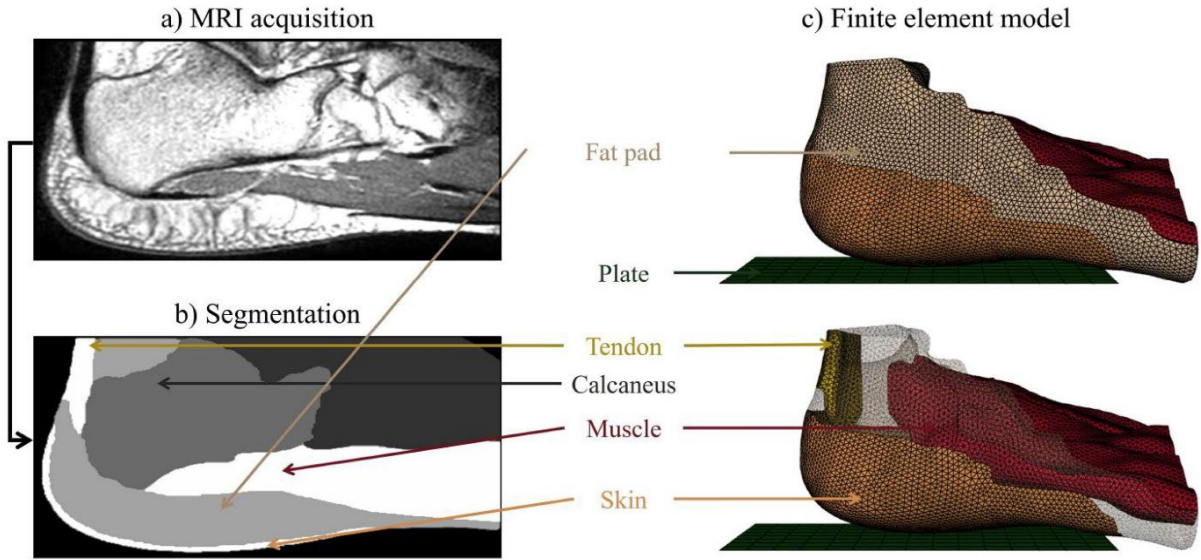

Figure 2

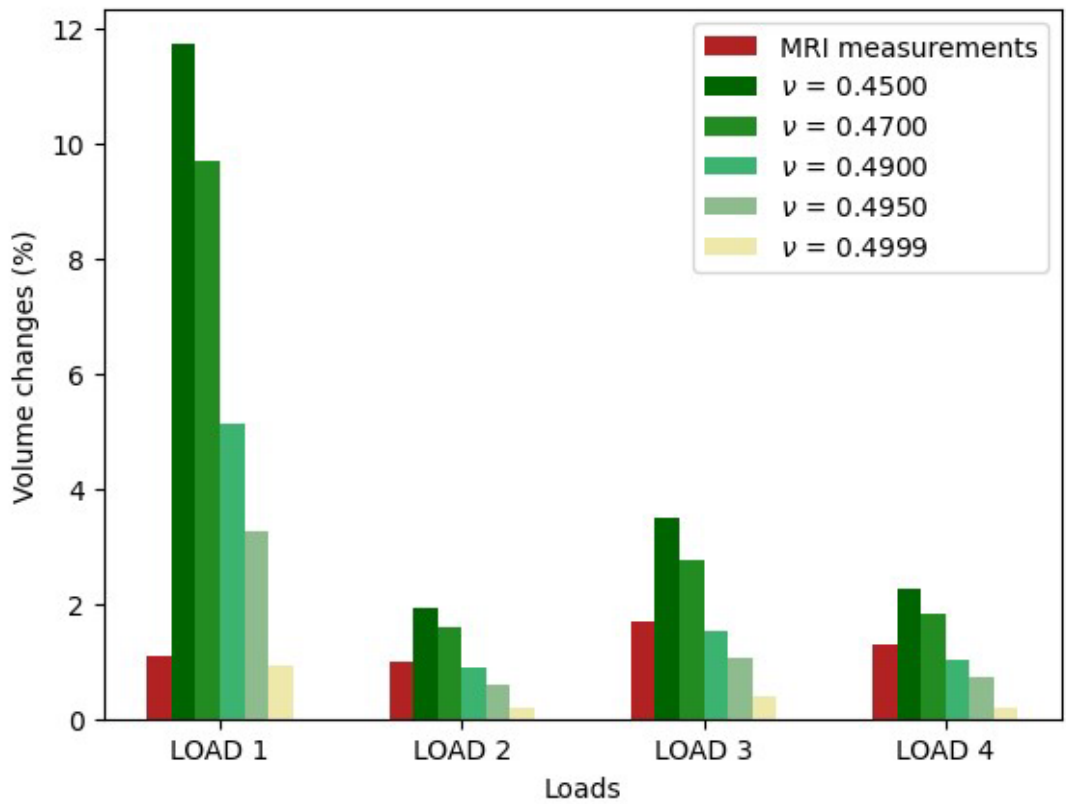

Figure 3

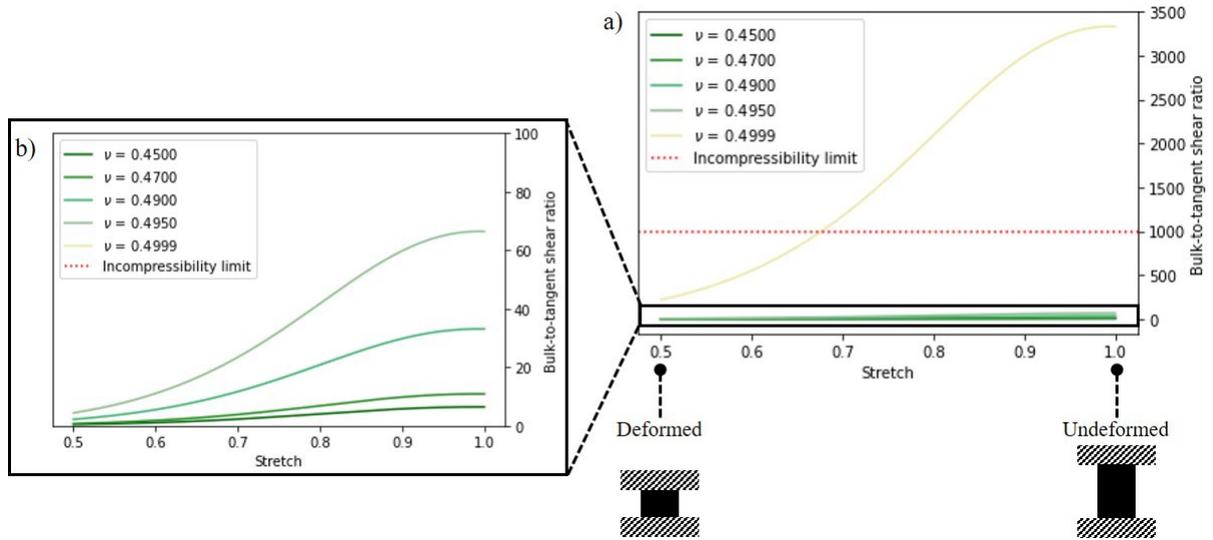

Figure 4

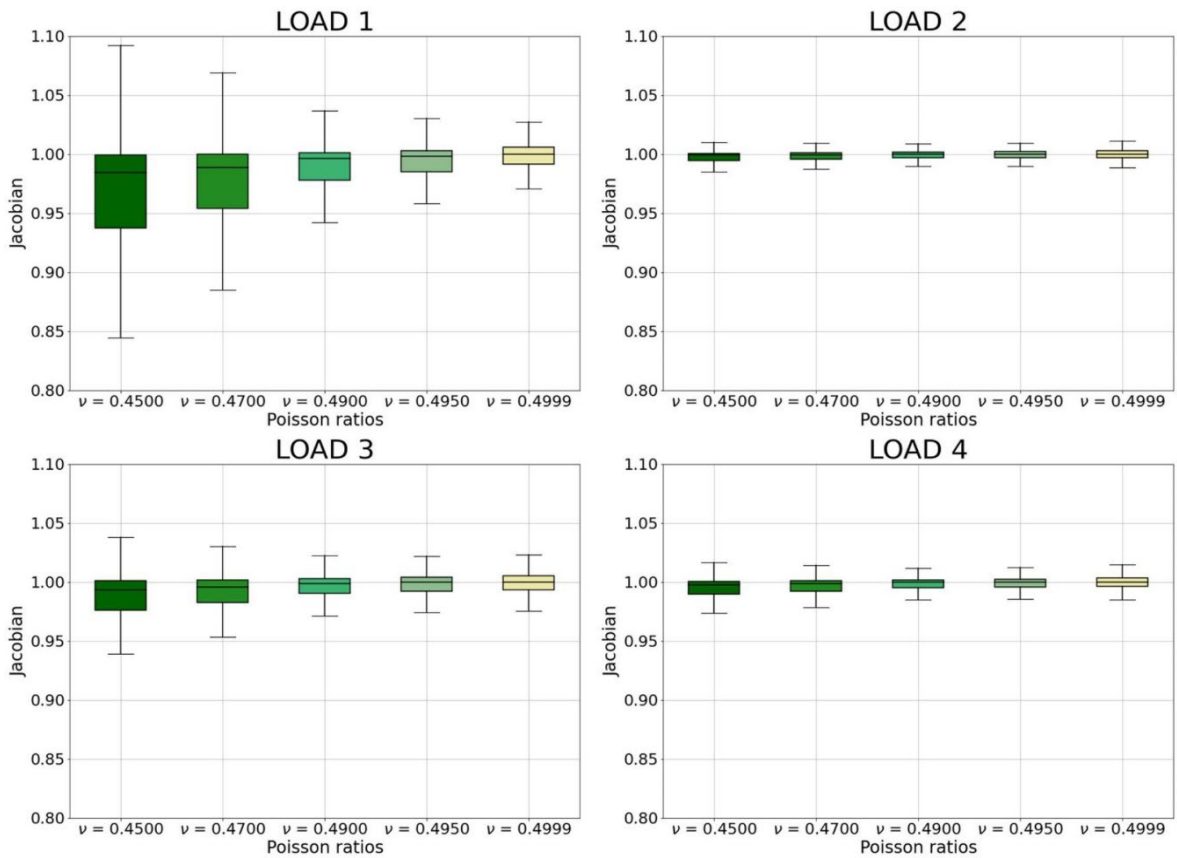

Figure 5

# 9. List of Tables

| Table 1 | Material properties of the soft tissues obtained from literature experimental data. |
|---|---|
| Table 2 | Values of Poisson ratio and equivalent bulk modulus and bulk-to-shear ratio. |
| Table 3 | Normal and shearing loads of all studied configurations and measured and computed plate displacement. US: horizontal displacement, UN: vertical displacement, FS: shearing load, FN: normal load. |
| Table 4 | Absolute values of maximal and mean volumetric strains. |

| Material parameters | µ (MPa) | α | $C_{10}$ (MPa) | $C_{20}$ (MPa) | $C_{30}$ (MPa) |
|---|---|---|---|---|---|
| Adipose tissues | 0.003 | 6.2 | - | - | - |
| Skin | - | - | 0.265 | 1.923 | - |
| Tendon | - | - | 9.654 | $1.897\ 10^2$ | $7.895\ 10^4$ |
| Muscle | - | - | 0.005 | 0.069 | 1.967 |

*Table 1: Material properties of the soft tissues obtained from literature experimental data.*

| Poisson ratio | 0.4999 | 0.4950 | 0.4900 | 0.4700 | 0.4500 |
|---|---|---|---|---|---|
| Bulk modulus (MPa) | 17.00 | 0.34 | 0.17 | 0.06 | 0.03 |
| Bulk-to-shear ratio | 5000 | 100 | 50 | 16 | 10 |
| Compressibility | 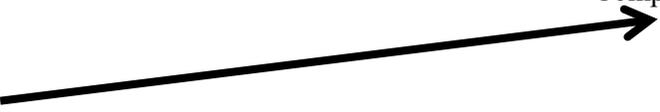 | | | | |

*Table 2: Values of Poisson ratio and equivalent bulk modulus and bulk-to-shear ratio.*

| Loading | LOAD 1 | | LOAD 2 | | LOAD 3 | | LOAD 4 | |
|---|---|---|---|---|---|---|---|---|
| *Force* | $F_S$ | $F_N$ | $F_S$ | $F_N$ | $F_S$ | $F_N$ | $F_S$ | $F_N$ |
| Experimental measures (N) | 0 | 140 | 0 | 15 | 45 | 15 | 12 | 15 |
| *Displacement* | $U_S$ | $U_N$ | $U_S$ | $U_N$ | $U_S$ | $U_N$ | $U_S$ | $U_N$ |
| Experimental measures (mm) | - | 8.3 | - | 5.5 | 6.7 | 5.5 | 4.1 | 5.5 |
| FEA, $\nu$ = 0.4500 | - | 13.4 | - | 7.0 | 5.4 | 6.6 | 1.6 | 6.9 |
| FEA, $\nu$ = 0.4700 | - | 12.7 | - | 6.9 | 5.2 | 6.6 | 1.6 | 6.8 |
| FEA, $\nu$ = 0.4900 | - | 11.0 | - | 6.6 | 4.9 | 6.3 | 1.5 | 6.6 |
| FEA, $\nu$ = 0.4950 | - | 10.2 | - | 6.6 | 4.8 | 6.3 | 1.5 | 6.5 |
| FEA, $\nu$ = 0.4999 | - | 9.3 | - | 6.6 | 4.6 | 6.3 | 1.4 | 6.5 |

*Table 3: Normal and shearing loads of all studied configurations and measured and computed plate displacement. $U_S$: horizontal displacement, $U_N$: vertical displacement, $F_S$: shearing load, $F_N$: normal load.*

| *Loads* | LOAD 1 | | LOAD 2 | | LOAD 3 | | LOAD 4 | |
|---|---|---|---|---|---|---|---|---|
| *Strains (%)* | *Max* | *Mean* | *Max* | *Mean* | *Max* | *Mean* | *Max* | *Mean* |
| FE model, $\nu$ = 0.4500 | 76 | 7 | 35 | 1 | 54 | 3 | 34 | 1 |
| FE model, $\nu$ = 0.4700 | 72 | 6 | 35 | 1 | 50 | 2 | 34 | 1 |
| FE model, $\nu$ = 0.4900 | 60 | 3 | 35 | 1 | 41 | 2 | 34 | 1 |
| FE model, $\nu$ = 0.4950 | 58 | 3 | 36 | 1 | 38 | 2 | 35 | 1 |
| FE model, $\nu$ = 0.4999 | 60 | 3 | 36 | 1 | 36 | 2 | 35 | 1 |

*Table 4: Absolute values of maximal and mean volumetric strains.*